\documentclass[fleqn,10pt]{wlscirep}
\usepackage[utf8]{inputenc}
\usepackage[dvipsnames]{xcolor}
\usepackage[T1]{fontenc}
\usepackage{subcaption}
\usepackage[export]{adjustbox}
\usepackage{wrapfig}
\usepackage[export]{adjustbox}
\usepackage{braket}
\usepackage[numbers]{natbib} 

\usepackage{multirow}
\usepackage{caption}
\usepackage{graphicx}
\usepackage{float}      
\usepackage{booktabs}   %

\title{Identifying preferred routes of sharing information on social networks}

\author[1,*]{Rozhin Mohammadikian}
\author[1]{Parsa Bigdeli}
\author[1]{Behrouz Askari}
\author[1,2,3,+]{G.Reza Jafari}
\affil[1]{Department of Physics, Shahid Beheshti University, Evin, Tehran 1983969411, Iran}
\affil[2]{Center for Communications Technology, London Metropolitan University, London N7 8DB, UK}
\affil[*]{rozhinmkian99@gmail.com}
\affil[+]{g\_jafari@sbu.ac.ir}

\begin{abstract}
The spread of information has become faster and wider than ever with the advent of social network platforms. The question raised in this study is whether information dissemination in social networks is random or follows a discernible structure. Our results from real-world hashtag data suggest that the spread of hashtags is not random and follows specific patterns. This study proposes two preferential models to explore how news spreads on social media. Specifically, we examine global and local preferential selection models and demonstrate that information dissemination aligns with these patterns. According to these two models, information flows are distributed through specific paths on networks. This suggests that new information tends to propagate along the same paths as previous news, with the specific pathways varying depending on the type of content. Finally, an examination of the propagation of political hashtags on Twitter confirms the existence of these paths that also emerge from the two preferential models.

\end{abstract}

\begin{document}

\flushbottom
\maketitle

\thispagestyle{empty}

\section{Introduction}

The flow of information between users influences not only their knowledge but also their decision-making processes and judgment. Today, a substantial portion of this information flow is carried out through online social networks, which have become an inseparable part of daily life. From social marketing~\cite{yang2010study} and misinformation mitigation~\cite{gausen2021can}, to predicting political elections~\cite{hu2017competing, lang2022opinion, lewis2005election} and issuing natural disaster alerts~\cite{kim2018social,zhang2016comprehensive}, understanding the mechanisms of information propagation in cyberspace has emerged as a central problem in multidisciplinary research.

Although users can theoretically connect with many others on contemporary online platforms, these connections are far from uniform. Pre-existing offline social structures, largely influenced by geographic constraints~\cite{grabowicz2014entangling, crandall2010inferring}, online visibility (e.g., celebrities, political figures or media organizations), and shared interests all contribute to heterogeneous interaction patterns. Moreover, connections are not used uniformly across topics. A person may be more likely to share content with certain contacts on specific subjects while engaging different contacts for other topics. This behavior gives rise to topic-dependent sharing preferences for each user, forming what is often referred to as an underlying diffusion network~\cite{poux2023dirichlet}. News propagation can then be conceptualized as a biased random walk on this topic-sensitive network~\cite{molaei2018information}. In this work, we investigate whether such route-level sharing preferences can emerge from simple generative mechanisms and whether these mechanisms can reproduce empirical patterns observed on social media platforms.

A large body of research has examined information diffusion in social networks~\cite{kurka2015online}, including theoretical modeling approaches~\cite{li2017survey}, empirical measurements of diffusion dynamics~\cite{PhysRevE.78.065102}, and computational or machine learning methods aimed at prediction and practical applications~\cite{firdaus2021retweet,bunyamin2016comparison}. Diffusion models take the underlying network as given and model a dynamical process unfolding on top of it, such as Threshold models~\cite{granovetter1978threshold,kempe2003maximizing,chen2009approximability}, Cascade models~\cite{goldenberg2001talk, kempe2003maximizing} and many others~\cite{singh2018survey}, each with a property best fit to simulate a particular spreading scenario in social networks. These frameworks focus on transmission events, temporal excitation, and cascade statistics, while typically assuming fixed edge weights or homogeneous transmission rules.

While diffusion models address the process of propagation, evolution models address the structure that constrains such propagation. For example, foundational models such as the Barabási–Albert (BA) model~\cite{barabasi1999emergence}
explain the emergence of hubs through a node-level reinforcement mechanism, i.e. the preferential attachment. Extensions such as the weighted growth model of Barrat, Barthélemy, and Vespignani (BBV)~\cite{barrat2004modeling} further incorporate edge weights, allowing link strengths to evolve as the network expands. These models belong to a class of structural evolution frameworks in which the network's growth is the primary object of study. Bridging the gap between purely dynamical models and purely evolutionary models are the co-evolving models of networks, such as the COEVOLVE~\cite{farajtabar2017coevolve} in which a diffusion process and link creation happens simultaneously.

Although these evolution models successfully capture node-level prominence, the literature reveals a notable gap in studies and modeling the emergence of link-level preferences and the formation and strengthening of specific ties through repeated interaction remains comparatively underexplored, which is one way to study route-level sharing preferences in social networks.

To address this gap, we propose two generative models of link-level preference formation: a global preferential model and a local preferential model. Both are inspired by the reinforcement principle underlying the BA model but extend it from node-level to link-level dynamics. While the BA mechanism produces recurrently attached nodes, our models also generate recurrently traversed links, leading to the emergence of preferred sharing paths.
In both models, probabilistic reinforcement at each time step gradually shapes an underlying weighted network that encodes user preferences. Information diffusion is then modeled as a biased random walk on this emergent network. To evaluate whether the models reproduce empirical sharing patterns, we employ two complementary preference measures: a modified weighted Jaccard index and a functional similarity metric. The aforementioned BBV model and a random (no-preference) model are taken as our benchmark and null models respectively. Applying our framework to political hashtag data from the X platform (formerly Twitter), we demonstrate that link-level reinforcement-based mechanisms can generate structured sharing patterns that are substantially closer to the empirical data, compared to the benchmark and the null models.

The remainder of the paper is organized as follows. Section 2 introduces the global and local preferential models. In Section 3, we describe the dataset and preprocessing steps, followed by introducing preference measures on networks. Section 4 reports the results of preference measures on both the empirical and simulation data, followed by conclusion.

\begin{figure}[t]
    \centering
    \includegraphics[width = \textwidth]{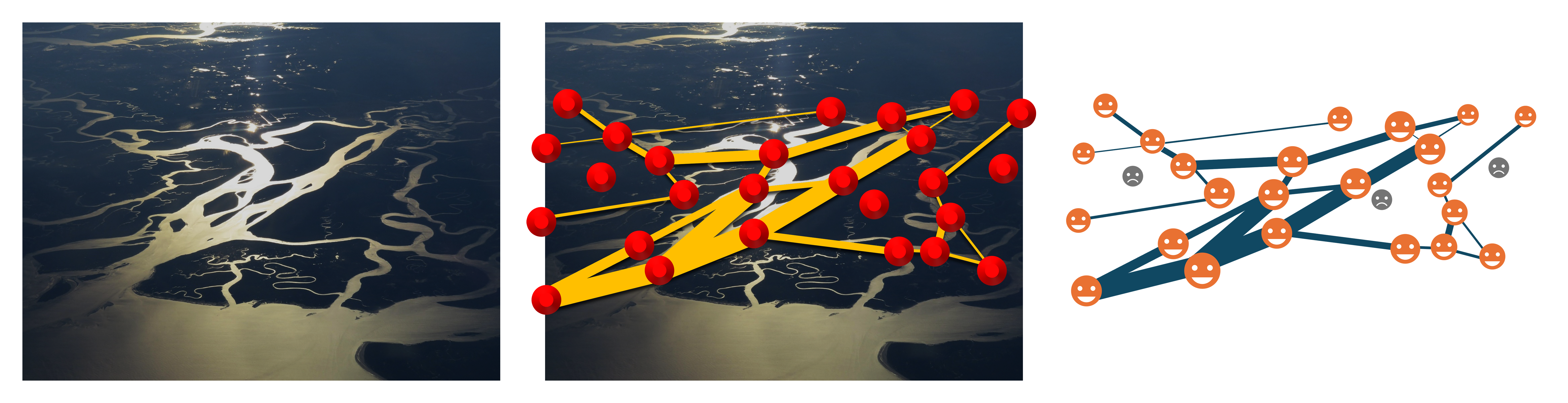}
    \caption{One case of preferred routes in nature is riverbeds~\cite{pixabay_swamp}, which form over time after many repetitions. The more water streams down a path, the more indented the riverbed will become. We take this idea to explain the formation of preferred routes of information sharing on social media.}
    \label{fig:preference1}
\end{figure}

\section{Mathematical Model} \label{sec:model}
To construct the model, we begin by examining empirical studies on the motivations and dynamics of social interactions. Rather than capturing the full behavioral complexity, we identify and incorporate key mechanisms that are most relevant for formal modeling.
Social and group living have long been recognized as a survival strategy among many species, including humans, providing advantages in reproduction, offspring care, foraging, and predator avoidance~\cite{rubenstein1978predation}. In the modern context, however, the motivations for human social interaction have evolved beyond basic survival and reflect higher-order psychological needs. This shift is particularly evident in the use of social media. Wong et al.~\cite{wong2017motivations} argue that content sharing on social networks is primarily driven by the desire to inform and entertain an audience but also stems from deeper motivations such as maintaining relationships, influencing others, asserting individuality, and affiliating with social groups. According to their findings, social media engagement is largely shaped by an intrinsic need to share.
They also mention that sharing content is influenced by three additional factors: (i) the recipient, (ii) the relationship between them, and (iii) the content, all of which we aim to incorporate into our model. The recipient and the strength of the relationship between the recipient and the sender determine the frequency—or equivalently, the probability—by which the content is shared. The strength of this relationship can arise from multiple causes, developing over time either internally through social media interactions or externally through offline connections.

We aim to model the former mechanism using our proposed framework, while treating the latter as a predefined input parameter, since external social connections often form long before an individual is eligible to use social media. Although these external relationships can evolve after joining a platform, we assume such changes are relatively slow and can be neglected in our dynamic formulation. Overall, the key point is that these relationships are both user-specific and time-dependent.
As highlighted in the third point above, the strength of a relationship can vary depending on the content being shared. According to Berger et al.~\cite{berger2012makes}, sharing practically useful information is motivated either by altruistic intentions, such as helping others, or by the desire to enhance one’s social image, for example, to appear knowledgeable to colleagues. Picone et al.~\cite{picone2020shares} reinterpret this latter motivation as a form of self-publication, drawing an analogy to journalism in which individuals prioritize the interests of their audience and selectively share topics aligned with those interests. This perspective supports the assumption that each social media user has context-dependent sharing preferences. For instance, a user may exhibit distinct sharing behaviors depending on whether the content is political, entertainment-related, or scientific. Consequently, the network of shared content can be decomposed into topical multiplexes that evolve independently and simultaneously. This concept is known as \textit{thematic multiplex}~\cite{hanteer2019innovative}.

Lastly, we also take into consideration the fact that some users are more prominent due to various reasons, such as being a celebrity or a political figure. Such prominence increases their visibility within the network and makes them more likely to be involved in social interactions~\cite{wasserman1994social}.
It is important to clarify that when referring to the roles of sender and recipient, we mean the temporal sequence in which individuals are exposed to a given piece of content. These roles can manifest differently across platforms. For instance, in messaging systems like email or direct messaging features of platforms such as Instagram, the sender possesses the content before the recipient and decides whether or not to share it. In contrast, on microblogging platforms such as X, the user who originally posts the content (e.g., the one being retweeted) is the sender, and the one who retweets it is the recipient, even though in this case, it is the recipient who actively decides to share it.

\subsection{Global Preference Evolution Dynamics} \label{ssec:global}
We consider a base network $G$, modeled as a weighted, undirected graph of size $N$, where nodes represent users. The degree $k$ of a node quantifies its prominence, and the edge weight $L$ between two nodes, an integer, represents the number of social interactions (termed \textit{events}) between them. Following the framework of Rabbani et al.~\cite{rabbani2021memory}, the dynamics proceed by selecting a sender node at random to initiate an event, capturing user activity. The sender is more likely to interact with a recipient of higher node degree, reflecting the increased prominence and hence visibility of such users. Once the recipient is chosen, the interaction is recorded by incrementing the corresponding edge weight by $x$, which is typically set to one when following the definition of the number of interactions, but it can also be set to other values to emphasize the influence of interactions. Accordingly, the probability of a node participating in an event, either as the sender or the recipient, at time $t_n$ is:
\begin{equation}
\label{eq:pi_ki_global}
\Pi_{k_i}(t_n) = \frac{1}{N} + \frac{k_i(t_n)}{\sum\limits_{i=1}^{N} k_i(t_n)}
\end{equation}
If $m$ nodes participate at each time step, the probability of event participation of node $i$ becomes $m\Pi_{k_i}$. In the continuum limit of time, the evolution of each node’s weighted degree follows:
\begin{equation}
\label{eq:dki/dt_global}
\frac{dk_i(t)}{dt} = \frac{m}{N} + \frac{mk_i(t)}{2mt+m_0}
\end{equation}
Here, $m_0$ denotes the total initial weights of the links, and $2mt$ accounts for the cumulative increase in total node degrees over time.
This dynamic is inspired by the Barab{\'a}si-Albert (BA) preferential attachment model but differs in some aspects. Unlike the BA model, our network maintains a fixed number of nodes.
Moreover, on the occurrence of each social event, it is the weight of the corresponding
link that changes explicitly, not the number of neighbors of the selected nodes, although the change in the weight of the links consequently
results in changing the weighted degrees of each node.

To compute the change in the weight of a link, we consider the probability that an event occurs between nodes $i$ and $j$, accounting for both cases in which either node may act as the sender and the other as the recipient. Assuming that each node is equally likely to initiate an event, the combined probability of interaction between them is given by the average of the two directional probabilities. Each of these probabilities is proportional to the prominence (i.e., degree) of the initiating node and normalized over the total degree in the system. Thus, the interaction probability becomes:
\begin{equation}
\setlength{\jot}{10pt}
\begin{split}
\Pi_{L_{ij}}(t)& = \frac{1}{2} \left( \frac{1}{N}\times\frac{k_i(t)}{(2mt+m_0)} + \frac{1}{N}\times\frac{k_j(t)}{(2mt+m_0)} \right) \\[5pt]
    &= \frac{(k_i(t)+k_j(t))}{2N(2mt+m_0)}
    \label{eq:P_Lij_gloabl}
\end{split}
\end{equation}
When $m$ interactions occur at each time step, this leads to the following rate of change for the link weight in the continuum limit:
\begin{equation}
\label{eq:dLij_dt_global_compact}
\frac{dL_{ij}}{dt} = \frac{m(k_i(t) + k_j(t))}{2N(2mt + m_0)}.
\end{equation}

This result is also consistent with the fact that $k_i = \sum\limits_{j=1}^{N}L_{ij}$ and indicates that the rate of increase in a link's weight is directly proportional to the average prominence of its two end nodes.

\subsection{Local Preference Evolution Dynamics} \label{ssec:local}
In the evolution dynamics of global preference, each node evaluates the prominence of all other nodes based on their weighted degrees, which are assumed to be globally visible across the network. This allows a node to initiate interaction with a highly prominent node even in the absence of prior connections. However, in many real-world settings, social interactions are influenced by shared history between individuals. To account for this, we introduce an alternative mechanism, which we refer to as the \textit{local preference} model.

While in the global preference model, a node’s decision to interact was based on the degrees of other nodes, in the local preference model, this decision depends on the existing link weights. If node $i$ is randomly selected as the sender at time step $t_n$, the probability of its interaction with node $j$ is proportional to the weight of their link, $L_{ij}(t_n)$, normalized by the sum of the weights of all links connected to node $i$, i.e., $L_{ij}(t_n)/k_i(t_n)$. Here, $k_i(t_n)$ denotes the weighted degree of node $i$, defined as the sum of the weights of its links to neighboring nodes.
To account for events initiated by either node $i$ or node $j$, we symmetrize the probability by averaging the two possibilities. The resulting marginal probability that a link is selected for an event is:

\begin{equation}
\label{eq:Pi_Lij_local}
\Pi_{L_{ij}}(t_n) = \frac{L_{ij}(t_n)}{2N}\left( \frac{1}{k_j(t_n)} + \frac{1}{k_i(t_n)} \right)
\end{equation}
Due to the nature of this dynamic, which dictates that the evolution of each link’s weight is proportional to itself, all links with zero weight will remain zero indefinitely. This is inconsistent with real-world social networks, where any two individuals can potentially connect, albeit mostly with low probability. To address this, we add a small constant $L_0$ to all possible links in the network, ensuring weak but nonzero interaction probabilities. This approach resembles the modification introduced by Price et al.~\cite{price1965networks}, a precursor to the BA model. We choose $L_0 \ll x$, where $x$ is the link weight increment (e.g., for $x = 1$, we set $L_0 = 0.01$). Consequently, the network becomes effectively fully connected. For notational simplicity, we redefine the adjusted link weights as $L_{ij}^{'} := L_{ij} + L_0$ for all $i,j \in G$. This leads to the following rate of change for the link weight in the continuum
limit:
\begin{equation}
\label{eq:dLij_dt_global_compact}
\frac{dL_{ij}^{'}}{dt} = \frac{L_{ij}^{'}(t_n)}{2N}\left( \frac{1}{k_j(t_n)} + \frac{1}{k_i(t_n)} \right).
\end{equation}
Accordingly, the probability of node $i$ being a participant of an event, either as the sender or the recipient, is:
\begin{equation}
\setlength{\jot}{10pt}
\begin{split}
    \label{eq:p_ki_local}
    \Pi_{k_i}(t_n) &= \frac{1}{2} \left( \frac{1}{N}\times1 + \frac{1}{N}\times\frac{L_{ij}(t_)}{k_j(t_n)} + \frac{1}{N}\times\frac{L_{il}(t_n)}{k_l(t_n)} + ... \right) \\
    &= \frac{1}{2} \left( \frac{1}{N} + \langle\frac{L_{ij}(t_n)}{k_j(t_n)}\rangle_{i} \right).
\end{split}
\end{equation}
The $\langle\frac{L_{ij}(t_n)}{k_j(t_n)}\rangle_{i}$ term refers to the average of $\frac{L_{ij}(t_n)}{k_j(t_n)}$ all $j$; naturally, those that are not a neighbor of $i$ will have a link weight of zero.  Eq. \ref{eq:p_ki_local} shows that in the local preference model, the probability of a node participating in an event as the recipient (the second term in parentheses) is determined by how favored the node is in the eyes of its neighbors, while the probability of being chosen as the sender (the first term in parentheses) remains the same for all nodes.

In the global preference model, all nodes determine their interaction preferences based on a common measurement, which is the degree of other nodes. As a result, node preferences are largely homogeneous across the network. The only source of variation arises from the constraint that self-loops are prohibited, meaning that no node can include itself in its own preference list. In contrast, the local preference model can introduce heterogeneity. Here, each node's preferences are shaped by the weights of its existing links, which is not necessarily the same as those of other nodes.

\begin{figure*}[t] 
    \center
    \begin{subfigure}[t]{0.4\textwidth}
    \includegraphics[height=5cm]{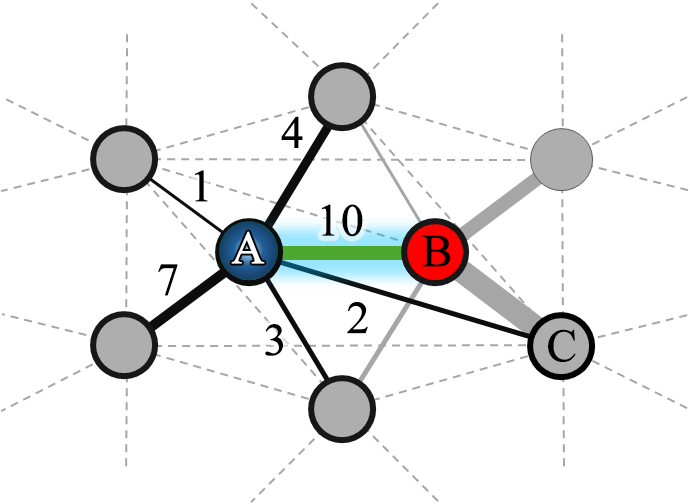}
    \label{fig:un-directed1}
    \caption{The probability of traversing the undirected link A-B (highlighted blue) when the sender is node A is approximately 37\%. Here, the preferred choice of node A is node B (colored green)} and not node C.
    \end{subfigure} %
    ~
    \hspace{0.04in}
    \begin{subfigure}[t]{0.4\textwidth}
    \includegraphics[height=5cm]{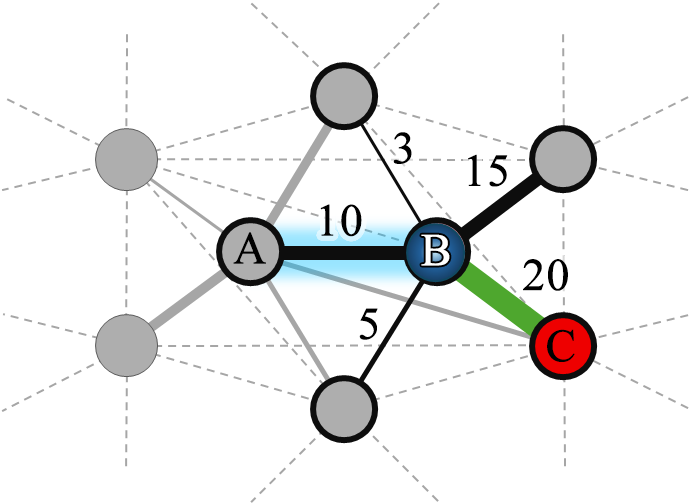}
    \label{fig:un-directed2}
    \caption{The probability of traversing the undirected link A-B (highlighted blue) when the sender is node B is approximately 19\%. Here, the preferred choice of node B is node C (colored green)} and not node A.
    \end{subfigure}
    \caption{A model with undirected network can account for an inherently directed phenomenon. For instance, the weight of the undirected link A-B can render different preferences depending on the sender node, breaking the directional symmetry of the role of an undirected link in news diffusion, while offering computational efficiency.
    \label{fig:un-directed}
    }
\end{figure*}

Also note that, for both models, the same link weight may lead to asymmetric preference decisions for the end nodes of the link; e.g., even if node B is the priority preference of node A, node A is not necessarily the priority preference of node B. This asymmetry in preference arises despite the underlying network being undirected and the adjacency matrix being symmetric. In practice, while the network topology is symmetric, the induced preference structure is not (see Fig.~\ref{fig:un-directed}). This observation justifies the use of an undirected adjacency matrix, which offers computational efficiency, even when modeling phenomena that exhibit inherently directed behavior.

\section{Route Preference In Information Flow} \label{Route Preference}
We propose that the evolution of social networks can follow a mechanism incorporating both the global and local dynamics introduced in Sections \ref{ssec:global} and \ref{ssec:local}.
However, it is not possible to directly verify this through comparing the networks that the models generate with the real-world underlying preference network for direct validation of this hypothesis, as the latter is not accessible to us;
at most, we can acquire a part of it (e.g. the follower-followee network, or friendship network).
But even this will not suffice; indeed, for most \textit{opinion formation} social interactions, social threads often involve people who don't follow each other ~\cite{kwak2010twitter}.
Furthermore, studies have shown that additional factors, including the writing style and sentiment of shared content, significantly influence interaction patterns~\cite{jimenez2021sentiments, stieglitz2013emotions}.
As a result, the follower-followee structure becomes a poor predictor of the likelihood of interaction. For example, on the X platform (formerly Twitter), only approximately 10\% of retweet interactions involving hashtags occur between users who are connected through a follower-follower relationship~\cite{bastos2012sticks}. Instead, what is readily observable is the flow of events between users, which we will acquire from the real-world social networks.

In conclusion, we assume that there is an effective underlying network, to which we do not have access, and that observable social events are spread by the preferences embedded within this network, analogous to the trajectory of a biased random walker on a weighted network.
If the underlying network is considerably preferential, then we expect the flow of information (corresponding to the walk of the random walker) to be carried out on preferred routes. As discussed in Section \ref{sec:model}, these dominant routes are expected to vary across contexts, but within each context, the pieces of information will most likely spread on the same routes, rather than diffusing randomly on the network.

\subsection{Real-World Data Retrieval}

As an example case, we used the hashtag retweet data from the Farsi X within political contexts\cite{mohammadikian2025retweet}. These data were acquired by Mohammadi et al.~\cite{mohammadi2022footprint} and further studied by Bigdeli et al.~\cite{bigdeli2024bots}, spanning from April 29, 2021, to June 24, 2021
(7 weeks prior to Iran's 2021 presidential election).
The trending hashtags were tracked daily, and 16 of them were used in this study. For inclusiveness and in order to minimize selection bias, the hashtags chosen consisted of those specific to each of the two opposing parties and also included neutral hashtags used by both groups.
To collect the data, a custom program called Twitter Machine was developed (named as such since at the time of data retrieval, X platform was officially called Twitter). Given a hashtag from the collected list and a start date, the program retrieved all tweets containing the specified hashtag via the Twitter Standard Search API and stored them in an SQLite database. The application parsed the JSON objects returned by the API, extracted the relevant metadata, and updated a user index maintained in a PostgreSQL relational database. While continuous automated requests were implemented to minimize data loss, completeness of retrieval is inherently subject to the technical constraints and rate limits of the Twitter Standard Search API. All users participating in election-related discussions identified by Twitter Machine were recorded in PostgreSQL to enable relational analyses. To ensure uninterrupted and efficient data collection, the program was deployed on a Virtual Private Server (VPS), allowing automated and continuous API requests within the platform’s rate limits. The collected data were securely transferred from the VPS to local storage infrastructure, where they were archived and processed for subsequent analysis.
Each tweet and retweet was saved along with the user ID of the retweeter and the user ID of the user being retweeted, along with the hashtags used in the body of the retweet and a timestamp of the retweet happening. The number of distinct retweets containing at least one of the election hashtags was 5701902, and they were posted by 140638 unique users.

A weighted, undirected retweet network was constructed from the data gathered for each hashtag, where nodes are the users and the link weights are the number of times a tweet containing that hashtag has been shared between two users, in either direction. This retweet network can be seen as the trajectory of a biased random walk on the underlying network of preferences discussed previously in this section. Note that the choice of an undirected retweet network will not pose any problem for our analysis, as discussed in Section \ref{ssec:local}. To further study the existence of preference, the two used measures will require us to perform preprocessing steps on the raw retweet data, the details of which are presented in Section~\ref{sssec:considerations}.

\subsection{Measures for Detecting Preference on Network}
To assess the presence of preference in the underlying network structure, we require quantitative measures to extract meaningful patterns from the observable information flow. Since the underlying preference network is inaccessible, we instead examine the structure of social interactions, such as information propagation events, as indirect indicators of these hidden preferences. Specifically, we employ two complementary metrics: (1) a modified weighted Jaccard index, which quantifies the overlap in communication paths while accounting for interaction weights, and (2) a functional similarity measure, which captures the consistency in the influence of nodes during information spread. Together, these measures allow us to infer whether information flows consistently along preferred routes, revealing latent structural biases in the network.

\subsubsection{Modified weighted Jaccard Index} \label{sssec:jaccard}
\begin{figure}[htbp]
    \centering
    \includegraphics[height=8cm]{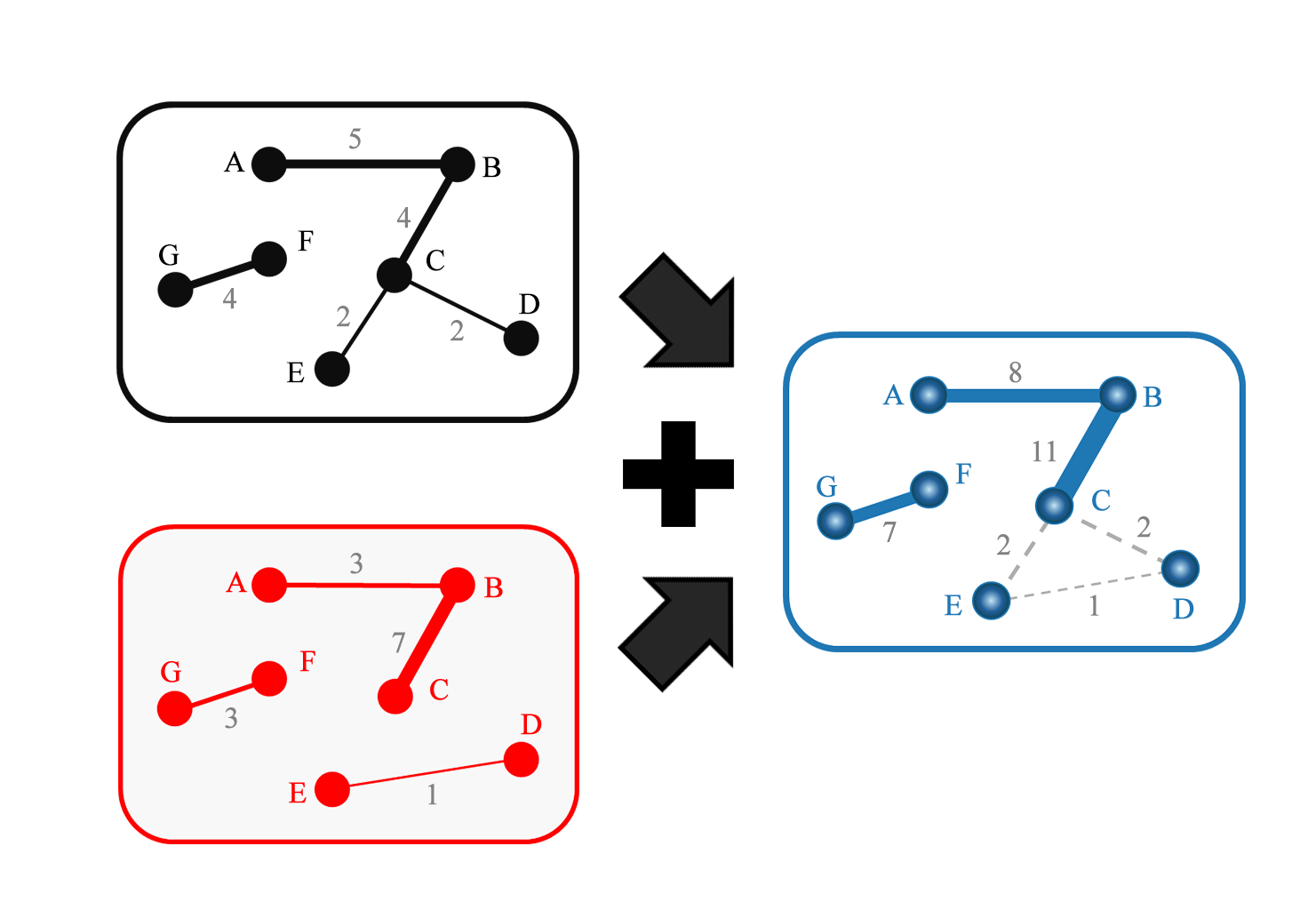}
    \caption{Overlap of retweet imprints of two hashtags: the red (lighter color) and the black (darker color) graphs on the left represent the sharing frequency of two distinct but contextually-similar hashtags between mutual users. The solid lines on the right show the mutual links used in the spread of both hashtags, while the dotted links represent those links used only for the sharing of one hashtag. For the example case of above, the modified weighted Jaccard index is calculated as $\tilde{J}_w(A,B) = \frac{8+7+11}{8+7+11+2+2+1}\approx0.84$.}
    \label{fig:overlap}
\end{figure}

Similar to the desire paths already introduced in the social sciences~\cite{nichols2014social}, we would like to study the overlap of the imprint of social interactions between social network users. If there exists preferred routes for the dissemination of news of a certain topic, then all hashtags relating to that topic will follow that route more or less. If the tweet is considered a biased random walker, the footprints of the walkers corresponding to each hashtag must have a considerable overlap with those of another hashtag.
Due to the inherent probabilistic nature of news dissemination, some randomness is inevitable, and non-preferred links will also be traversed. However, what is critical is the relative frequency with which preferred links are traversed as opposed to non-preferred ones.
To identify and compare the preferred dissemination routes for a given topic, we quantify how often each link associated with a hashtag is traversed (i.e., the number of times news is shared through that link), as illustrated in Fig.~\ref{fig:overlap}.

The typical [unweighted] Jaccard index gives the ratio of the overlap of two sets to the union of them (i.e. $J(A, B)=|A\cap B|/|A\cup B|$), which has already been a staple in studying social networks\cite{evkoski2021community,jafariasbagh2014clustering}. Naturally, no repetition is considered if we want to calculate the Jaccard index of two multisets (a.k.a. sets but with arbitrary repetition of elements). Therefore, the unweighted Jaccard index will only take into account the links traversed without considering the number of repetitions on that link. The weighted Jaccard index~\cite{costa2021further} is a solution to measuring the overlap of multisets, where the frequency of the news sharing between each pair of users is taken into account. In order to embed the intuition behind the desire paths and the depth of footprints that retweets leave behind, we use a modification of this measure, which essentially gauges the percentage of all the steps taken in the sharing of two hashtags that have traversed mutual links. For two hashtags with adjacency matrices $\mathbf{M_1}$ and $\mathbf{M_2}$, this could be calculated as follows:
\begin{equation}
\label{eq:mod_jaccard}
\tilde{J}_w(\mathbf{M_1},\mathbf{M_2}) = \frac{\sum\limits_{i,j}\left[\left(({M_1}_{,ij}+{M_2}_{,ij} \right)H({M_1}_{,ij} )H({{M_2}_{,ij}})\right]}{\sum\limits_{i,j} \left[ {M_1}_{,ij}+{M_2}_{,ij} \right] }
\end{equation}
where $H(x)$ is the Heaviside function and ${M_1}_{,ij}$ and ${M_2}_{,ij}$ are the $ij$th component of $\mathbf{M_1}$ and $\mathbf{M_2}$ respectively. If these matrices are not of the same dimensions, which is often the case with real-world data, the data must undergo a pre-processing step, discussed in Section \ref{sssec:considerations}.
\vspace{-3mm}

\subsubsection{Functional Similarity} \label{sssec:similarity}
To further investigate the presence of preference within the underlying networks of contextually similar hashtags, we focus on the behavior of individual nodes when exposed to messages of related topics. That is, we aim to determine which nodes a given node is most likely to share a received message with.
A suitable metric for this purpose is the cosine similarity~\cite{newman2004detecting,newman2012communities}. Generally, the cosine similarity between two normalized vectors is given by their inner product, which quantifies the degree of alignment between the vectors in their respective vector space. When these vectors represent the sharing behavior of a specific node across two different networks, the cosine similarity reflects how similarly that node functions in those networks, hence also called the \textit{functional similarity}.
For the weighted, undirected network corresponding to hashtag $m_1$, represented by adjacency matrix $\mathbf{M_1}$, the retweet state vector of node $i$ is constructed as described in Eq.~\ref{eq:similarity_vector}, and normalized by the norm of the space (inner product),
\begin{equation}
\label{eq:similarity_vector}
\Ket{p^{(m_1)}_i} = \frac{1}{\sqrt{\sum\limits_{j=1}^{N} {M_1}_{,ij}^2 }}
\begin{pmatrix}
{M_1}_{,i1}\\ {M_1}_{,i2}\\ \vdots \\ {M_1}_{,iN}
\end{pmatrix}
\end{equation}
where ${M_1}_{,ij}$ shows the frequency of retweets between node $i$ and node $j$ in hashtag $m_1$. The functional similarity of a node in the dissemination of two hashtags $m_1$ and $m_2$ with adjacency matrices $\mathbf{M_1}$ and $\mathbf{M_2}$ is then determined by:
\begin{equation}
\label{eq:functional_similarity}
S_{i}^{m_1,m_2} = \Braket{p^{(m_1)}_i | p^{(m_2)}_i}  = \frac{\sum\limits_{j=0}^N {M_1}_{,ij}{M_2}_{,ij}}{\sqrt{\sum\limits_{j=1}^{N} {M_1}_{,ij}^2 \sum\limits_{j=1}^{N} {M_2}_{,ij}^2 }}
\end{equation}
Averaging over functional similarities of multiple nodes $\langle S_{i}^{m_1,m_2} \rangle$ results in an overall measurement of the existence of preference. Note that the dimension of the state vector $\ket{p^{(m_1)}_i}$ is by default the size (number of users) of the retweet network of hashtag $m_1$. However, most of the time modifications must be made for Eq. \ref{eq:functional_similarity} to be applicable, which we will discuss in Section \ref{sssec:considerations}. Additionally, the choice of whether to pick the $i$th row of an adjacency matrix or its $i$th column to construct the state vectors used in Eq. \ref{eq:functional_similarity} is naturally irrelevant for undirected networks, such as the case of this study.

\begin{figure}[htbp]
    \centering
    \begin{minipage}[t][0.3\textheight][t]{0.4\textwidth}
        \centering
        \vspace{2mm}
        \includegraphics[width=\textwidth, valign=t]{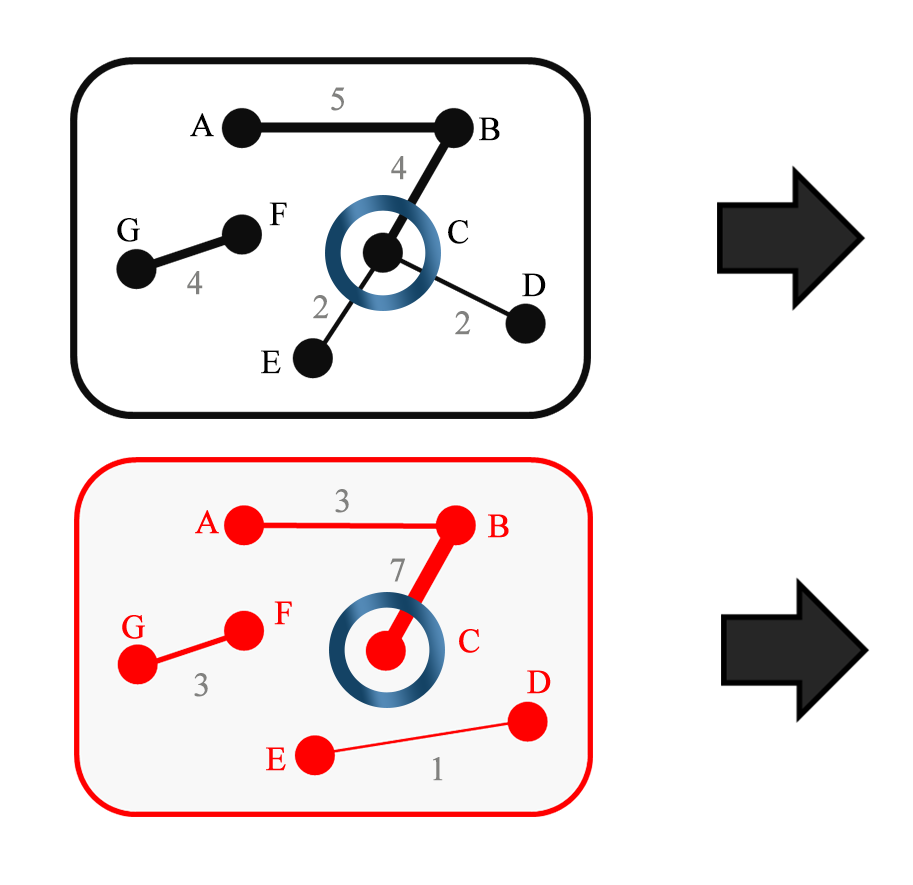}
    \end{minipage}
    \begin{minipage}[t][0.3\textheight][t]{0.3\textwidth}
        \centering
        \begin{equation*}
            \Ket{p^{(black)}_C} = \frac{1}{\sqrt{24}}
            \begin{pmatrix}
           0 \\ 4\\ 0\\ 2\\ 2 \\0\\0
            \end{pmatrix}
        \end{equation*}
        \begin{equation*}
            \Ket{p^{(red)}_C} \; =\; \frac{1}{\sqrt{49}} \;
            \begin{pmatrix}
            0\\7\\ 0\\ 0 \\0 \\0 \\0
            \end{pmatrix}
        \end{equation*}
        \vfill
    \end{minipage}
    \caption{Retweet imprints of a single node in two hashtags: In sharing two different hashtags, a node of interest (marked with gradient color) can act with partial similarity, calculated by Eq. \ref{eq:functional_similarity}. The functional similarity of the node $C$ in the above example will be $S_{C}^{black,red} = \braket{p^{(black)}_C | p^{(red)}_C} \approx 0.82$. One way of measuring the overall similarity of two hashtags is to take the average of the functional similarities of all their nodes.}
    \label{fig:figure_math_combined}
\end{figure}
Similar measures based on inner product have been used for the study of retweet networks, such as \textit{favoritism} of nodes~\cite{kwak2010twitter,almaas2004global}. However, favoritism is measured for a single node over one set of retweets, whereas functional similarity studies a single node over two sets of retweets (the difference might be analogized to the difference between autocorrelation and cross-correlation of series). The closer the functional similarity of a node is to unity over two hashtags, the more preferential the underlying network is.

\subsubsection{Considerations}
\label{sssec:considerations}

To apply these measures of preference to real-world data, we should make two considerations:

\begin{enumerate}
\item
Two hashtag datasets collected over similar time spans will rarely produce retweet network adjacency matrices of the same size; that is, the number of users involved in the sharing of each hashtag can differ significantly. However, to apply the modified weighted Jaccard index and functional similarity measures (Eqs. \ref{eq:mod_jaccard} and \ref{eq:functional_similarity}), the underlying networks must be of equal size. To assimilate the dimensions of two retweet networks, we can either expand their nodes to include the union of their users by adding isolated nodes or shrink them to sub-networks that include only the intersection of their users. The only difference is that the former will have additional zero terms in the summations of these measurements. The latter will eliminate the users who had only participated in the propagation of one of the two hashtags. In fact this makes our study more sensible, since our goal is to find whether a user will share similar topics from the same old route or not. So the user should participate in sharing both of them in the first place before we can even begin to study the overlap or similarity of the propagation routes.

\item
Discrepancy between the total number of retweets (i.e. the total number of walks of the biased random walker on the underlying network) of two hashtags can alter the results obtained from Eq. \ref{eq:mod_jaccard}.
To avoid this unwanted effect, we normalize the weights of all the links (i.e. the components of the adjacency matrix) to the total number of weights. This ensures that variations in $\tilde{J}_w$ only reflect the strength and the structure of sharing preferences between nodes rather than their overall activities. For functional similarity, no such normalization is required beyond what is already performed in Eq. \ref{eq:similarity_vector}, as the state vectors are already normalized.
\end{enumerate}

\subsection{Simulation}
\label{ssec:simulation}
Before performing the simulation, we note that some hashtags contain mutual retweets, meaning individual retweets may include multiple hashtags simultaneously. Including these retweets in the analysis of preference introduces sampling bias, as such retweets artificially inflate the similarity measures due to repetition. To prevent this bias, we divide each hashtag's retweet data into two temporal halves (the first and second halves of the 7-week period) and compare each pair of hashtags using data from different halves. This approach doubles the number of compared pairs from 120 to 240, with two data pairs for each hashtag pair corresponding to the differing temporal halves.

For the simulation, two initial underlying networks of $N=400$ were evolved under two rules: the global preference dynamic (Section \ref{ssec:global}) and the local preference dynamic (Section \ref{ssec:local}). For the initial network of the global preference model, $m_0=5$ links with weights of unity were randomly added to a weighted, undirected null network, while for the local preference model, the initial network was a fully connected network with link weights of unity. These two networks evolved independently to generate the underlying network on which we later simulate biased random walks, corresponding to the propagation of hashtags. We selected an arbitrary evolution time for each case, guided primarily by the need to obtain meaningful results in the subsequent analysis. We chose $t=300$ for the global preference model and $t=500$ for the local preference model. At each timestep, $m=3$ links were chosen to be increased by $x=10$ units. The choice of the parameter $x$ is relative to the initial weight of the links.

To benchmark the proposed global and local preference models, we implement a third network evolution model introduced by Barrat, Barthélemy, and Vespignani (BBV) \cite{barrat2004modeling}. In this model, new nodes attach preferentially to existing nodes according to their weighted degrees (or their \textit{strengths}), which also update the existing link weights proportional to the current interaction intensity. We use the standard formulation described in~\cite{barrat2004modeling}, with simulation parameters of $N=400$, $m=3$ to match the configuration of the global and local preference simulations, and also $\delta=6$ and $w_0=4$. The choice of $\delta$ and $w_0$ was made so the receiver node's weighted degree increases similar to the simulations of the global and local preference models (in other words, $\delta + w_0 = x = 10$, according to the BBV model. However the difference is that in the BBV model, the sender's weighted degree would only increase $w_0$ units, unlike the global and local preference models, which is one of its inherent properties.

As a null hypothesis representing the absence of sharing preferences, we also considered a fourth underlying network, fully connected, with homogeneous link weights. In this network, no node has any preferential priority over others.

With an adiabatic assumption that the changes in the preferences of the underlying networks happen on a greater timescale than the typical time between the rise and decay of hashtags~\cite{glasgow2013hashtag}, we assume the changes in individual preferences caused by the sharing of news of specific single hashtags (for example, in a time span of 7 weeks)
are negligible compared to the overall time span of users' presence, which has a median of 3-5 years~\cite{api}. With that in mind, we simulated hashtag retweets on the mentioned four networks without the retweets affecting the preferences of nodes (i.e., the weight of the links are not changed). The retweet simulation is essentially a biased random walk starting from an initial node and spreading based on each node's preferences. In other words, the probability $p$ of the random walker jumping to a neighboring node is proportional to the weight of the links of the node it is on.

Another assumption discussed in Section \ref{sec:model} is that hashtags with similar topics disseminate on the same underlying network. Therefore, we repeated the simulation of the biased random walks on each of the four underlying networks for a particular number of times (128 times, chosen so that 8 repetitions accounted for the 16 corresponding hashtags).
For each repetition, we added noise to the jumping probabilities from each node. We obtained the noisy jumping probabilities with a noise intensity $\eta$ from a simple mapping:
\begin{equation}
\label{eq:noise}
f: [0,1] \to [0,1], \quad \forall \eta\in[0,1]: \quad f(p) = p(1-\eta) + \frac{1}{N-1}\eta
\end{equation}
This mapping is such that for $\eta=0$ the probabilities of jumping are strictly the same as the probabilities induced by the underlying network, and for $\eta=1$, the probabilities are strictly equal with no preference. Also note that the $N-1$ in the denominator accounts for a possibility of traversing to all the nodes of the network except the one the walker is standing on.
For each of the simulation repetitions, $\eta$ was picked from a Gaussian distribution with a mean of $0.3$ and a standard deviation of $0.15$, where the values were bounded to the interval of $[0,1]$.

To better mimic the real-world, we restricted the initiating nodes of the random walks, meaning that only a certain subset of nodes started the chain of retweets by tweeting an original tweet. A quarter of the nodes were chosen as the initiating subset of the network, corresponding to the 25\% of the users of X who publish 95\% of the original tweets on this platform~\cite{mcclain2021behaviors}.
Finally, we used a different number of total steps for the random walk of each hashtag.
The number of steps for each simulation ensemble
was chosen so that the distribution of the ratio of total retweets to the number of participating users of the hashtags of the real
world data was satisfied.

After performing the retweet simulation as such, it is time to apply the preference existence measures mentioned in Sections \ref{sssec:jaccard} and \ref{sssec:similarity} on each pair of the hashtags both for the empirical data and their simulated counterparts separately.

\section{Results and Discussion}
\label{sec:results}

\begin{figure}[htb]
  \centering
  \begin{subfigure}[t]{0.47\textwidth}
    \centering
    \includegraphics[width=\linewidth]{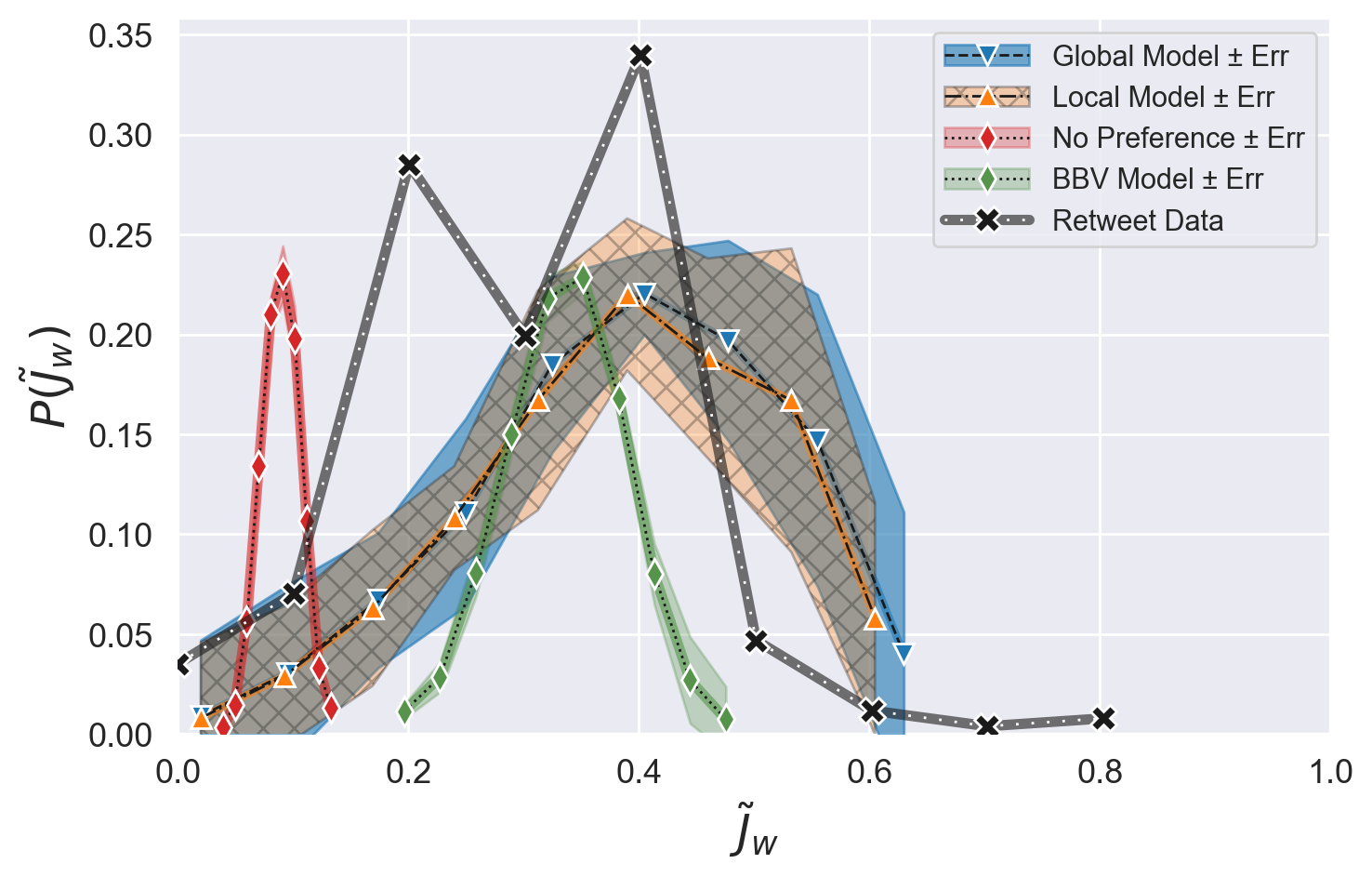}
    \caption{Probability distribution of modified Jaccard index of pairs of hashtag retweets in data and simulations.}
    \label{fig:fig1}
  \end{subfigure}%
  \hspace{0.04in}
  \begin{subfigure}[t]{0.47\textwidth}
    \centering
    \includegraphics[width=\linewidth]{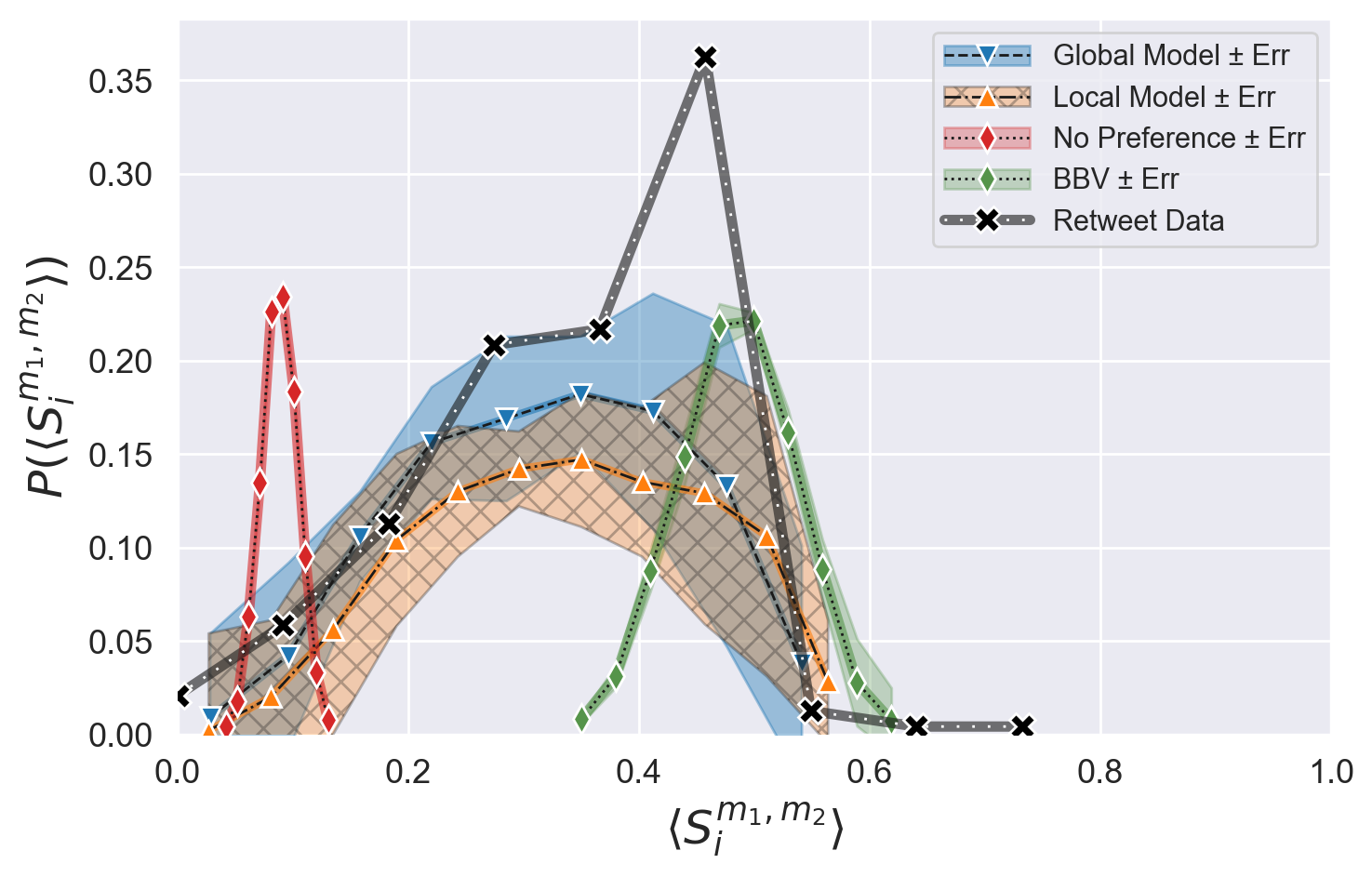}
    \caption{Probability distribution of mean functional similarities of nodes in pairs of hashtag retweets in data and simulations.}
    \label{fig:fig2}
  \end{subfigure}
  \caption{Results of measures of the existence of preference in the dissemination of tweets. The No Preference and the BBV models act as the null model and benchmark model respectively.}
  \label{fig:results}
\end{figure}

The results of the modified weighted Jaccard index and mean functional similarity of nodes are depicted in Fig. \ref{fig:results}. This result shows the probability distribution of each measure when applied to all the available pairs of networks in the case of the empirical data, along with three cases of retweet simulations conducted on the four underlying networks of Section \ref{ssec:simulation}. We used Sturge's rule to determine the number of bins in the figures.

The sampling errors corresponding to the four simulation cases are shown with colored areas (where the errors corresponding to the BBV and No Preference are so low that they are hardly visible).

The deficiencies in the distributions of the proposed models against the retweet data show themselves in the p-values obtained by the Kolmogorov-Smirnov (KS) test (Table \ref{tab:ks_test}), calculated between the retweet data and the retweet simulation on the four models. This lack of exact fit to the data can be attributed to the two models' abstraction of the phenomena of information sharing, where it fails to encompass the full complexity of this social-technological process, such as the presence of modular structures in the network of users~\cite{newman2006modularity}, the heterogeneity of activity rates for the sender nodes~\cite{ying2018user,vaca2014modeling}, or even exogenous factors affecting the retweet tendencies~\cite{bakshy2012role}.

However, as is seen in Fig. \ref{fig:results}, both the model and the retweet data span a range of values that the BBV and No Preference models cannot acquire in both the Jaccard index and functional similarity. This shows that, firstly, indeed there exist sharing preferences among users in this example social network case which the No Preference model cannot capture for obvious reasons, and secondly, the mechanism that generates the sharing preference in our proposed models works better at containing the spectrum of the results in the empirical data than the benchmark model, the BBV. Comparison of distributions in Fig. \ref{fig:fig1} and \ref{fig:fig2} is quantified in Fig. \ref{fig:JSD}, where the Jensen-Shannon Divergence (JSD)~\cite{lin2002divergence} was calculated between the empirical retweet data and the four models' retweet simulations, which measures how distinguishable two distributions are by averaging how far each distribution is from their shared midpoint distribution. This measure is bounded between zero (distributions are exactly the same) and $ln2\approx 0.693$ (distributions are completely different). The JSD reports an overall information difference between two distributions, whereas the KS statistics indicates the largest local discrepancy between them. Thus for a quantitative comparison of two distributions which were not expected to fit exactly as discussed before, the JSD is a more descriptive measure.

It is worth noting that while the modified Jaccard index relies on all steps of the sharing of one hashtag and how all the nodes involved have behaved, the functional similarity is affected only by each individual node's preference, regardless of how the introduced stochasticity in the retweet simulation along with the topology of the underlying network might affect the whole path, thus rendering a better measure of the existence of individual preferences. In Fig. \ref{fig:JSD}, it is seen that the modified Jaccard index is closer to the real-life data than its mean functional similarity measure for the BBV model, considering the errors. In contrast the global and local preference models yield reverse results, showing that their mechanisms have been more successful in producing individual sharing preferences such as the real life social networks.

\section{Conclusion}
\label{sec:Conclusion}
In this study, we found that the spreading of news on social networks is not random and that there are specific preferential paths for this news flow. Knowing these paths will play an important role in managing or limiting its dissemination. We introduced two mathematical models of global and local preferences to evolve networks with no preference into networks that exhibit preferential behaviors similar to those of a real-world case (e.g., retweets of hashtags on the X social platform). We measured preferential behavior using a modified weighted Jaccard index and mean functional similarity. A sample of retweet data of trending hashtags confirmed the effectiveness of these two models and the preferential dissemination of news on social networks, compared to a benchmark and a null model. The results of the simulated models, although they differ in details from the real-world data, capture the essence of the existence of preference in news dissemination seen in real-world data. This suggests that the mechanism of the birth of preferences for social users can follow the dynamics of the introduced models. It is important to emphasize that these preferred transmission paths are dependent on the type of news. A user's choice of recipient may vary based on the content; for example, one contact may be preferred for political news, while another is favored for humorous content.

Future works may expand or combine global and local preferential models to better resemble a case of real-world interest. The choice of the evolution time of the underlying network is another case that can be studied further and can potentially give insight into the age of the underlying network, which directly connects to temporal analysis over larger time frames. It is also possible to study how the underlying network of two hashtags with considerably different topics differs from one another using the same measures of preference existence.

\begin{minipage}[t]{0.48\textwidth}
\vspace{0pt}
\centering
\renewcommand{\arraystretch}{1.2}
\setlength{\tabcolsep}{5pt}

\begin{tabular}{|l|cc|}
\hline
\multicolumn{3}{|c|}{\textbf{Modified Jaccard Index}} \\ \hline
\textbf{Simulated Model} & \textbf{Statistic} & \textbf{p-value} \\
\hline
Global        & 0.469 & $< 10^{-15}$ \\ \hline
Local         & 0.156 & $5.547\times 10^{-5}$ \\ \hline
BBV           & 0.142 & $7.086\times 10^{5}$ \\ \hline
No Preference & 0.913 & $< 10^{-15}$ \\
\hline
\multicolumn{3}{|c|}{\textbf{Mean Functional Similarity}} \\ \hline
\textbf{Simulated Model} & \textbf{Statistic} & \textbf{p-value} \\
\hline
Global        & 0.179 & $1.96\times 10^{-6}$ \\ \hline
Local         & 0.298 & $< 10^{-15}$ \\ \hline
BBV           & 0.350 & $< 10^{-15}$ \\ \hline
No Preference & 0.922 & $< 10^{-15}$ \\
\hline
\end{tabular}
\captionof{table}{KS statistics and corresponding p-values of the comparison of the empirical data from X vs. the retweet simulation based on the two proposed models (global and local), the benchmark model (BBV) and the null model (No Preference), calculated for the two preference measures. P-values less than the machine's precision are presented by $<10^{-15}$.}
\label{tab:ks_test}
\end{minipage}
\hfill
\begin{minipage}[t]{0.48\textwidth}
\vspace{0pt}
\centering
\includegraphics[width=\textwidth]{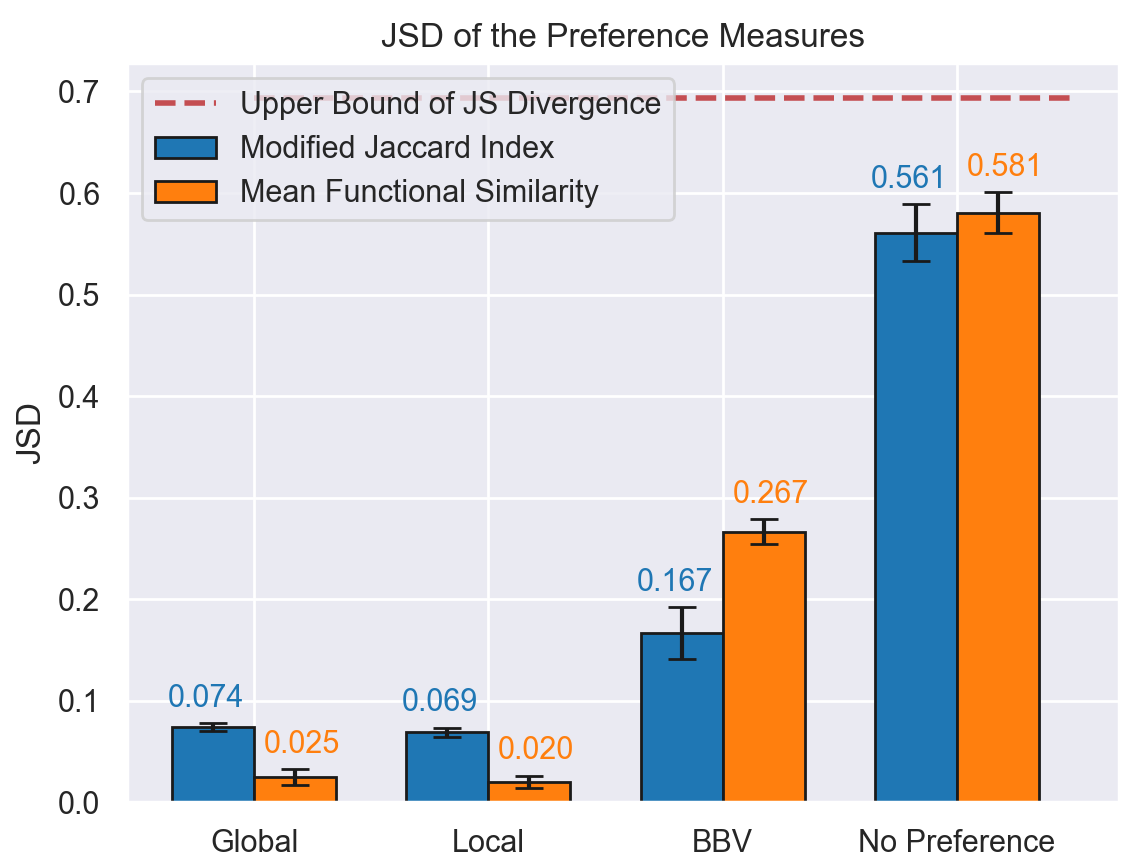}
\captionof{figure}{JSD for the results of the preference measures between the empirical data from X and the four simulated models. The JSD measure is bounded in the $[0,\ln 2]$ interval, with zero indicating perfect fit between two distributions and $\ln 2$ indicating maximal difference.}
\label{fig:JSD}
\end{minipage}

\section*{Acknowledgment}
We would like to thank Parham Moradi for his valuable assistance with data collection.

\bibliography{Bibliography}

\section*{Author contributions statement}
R.M. contributed to the conception of the project, conducted data analysis and simulations, and wrote the manuscript. B.A. contributed to project conception and manuscript writing. P.B. contributed to project conception and data handling. R.J. contributed to project conception, manuscript writing, and supervised the overall project.

\section*{Additional information}

\textbf{Competing interests} \\
The authors declare no competing interests.

\vspace{1em}

\noindent \textbf{Data availability} \\
The simulation code used in this study is available at \href{https://github.com/rozhinmkian/code-for-social-preferred-routes-2025}{the project's GitHub repository}. The datasets supporting this article have been uploaded on Dryad\cite{mohammadikian2025retweet}.

\vspace{1em}

\end{document}